\documentclass[%
 reprint,
 amsmath,amssymb,
 aps,
]{revtex4-1}

\usepackage{graphicx}% Include figure files
\usepackage{dcolumn}% Align table columns on decimal point
\usepackage{bm}% bold math
\begin{document}
\def\k #1{\bm{#1}}

\preprint{APS/123-QED}

\title{Triplet excitations in graphene-based systems}% Force line breaks with \\

\author{Vladimir Posvyanskiy}
\email{posvyanskij@fys.ku.dk}%Lines break automatically or can be forced with \\
\author{Logi Arnarson}%
\author{Per Hedeg{\aa}rd}%
\affiliation{%
Niels Bohr Institute, University of Copenhagen, Universitetsparken  5, 2100 Copenhagen, Denmark
}%

\begin{abstract}
In this article we investigate the excitations in a single graphene layer and in a single-walled carbon nanotube, i.e. the spectrum of magnetic excitations is calculated.
In the absence of interactions in these systems there is a unique gap in the electron-hole continuum. We show that in the presence of Coulomb correlations new states, magnons, appear in this forbidden region. Coulomb interaction is examined in the context of Pariser-Parr-Pople (PPP) model which takes into account long range nature of interaction. The energy of new bound states depends on the strength of Coulomb forces.
The calculations are performed for arbitrary electron-hole ($e-h$) momentum $\k q$ what allows to find the  magnons dispersion law $\varepsilon(\k q)$ , effective mass $m^*$ and velocity $v_{gr}$.
Finally, we determine the critical values of system parameters when this type of excitations can exist.
%\begin{description}
%\end{description}
\end{abstract}

\pacs{Valid PACS appear here}% PACS, the Physics and Astronomy
\maketitle

%\tableofcontents

\def\dsp {\downarrow}
\def\usp {\uparrow}
\def\Bdone #1#2{\left(B^{#1,#2}_{+}\right)^{\dag}}
\def\Bdmone #1#2{\left(B^{#1,#2}_{-}\right)^{\dag}}
\def\Bdzero #1#2{\left(B^{#1,#2}_{0}\right)^{\dag}}
\def\Adzero #1#2{\left(A^{#1,#2}_{0}\right)^{\dag}}
\def\Bone #1#2{B^{#1,#2}_{+}}
\def\Bmone #1#2{B^{#1,#2}_{-}}
\def\Bzero #1#2{B^{#1,#2}_{0}}
\def\Azero #1#2{A^{#1,#2}_{0}}

\section{Introduction}
Since the first isolation of graphene \cite{isolation}, a single two-dimensional (2D) atomic layer of graphite, it has attracted a lot of attention of both experimentalists and theoreticians. Such characteristics as high conductivity makes graphene a candidate for variety of modern nanoelectronics applications.
Mainly, graphene differs from usual 2D materials in that electrons have a linear relativistic like dispersion law and zero band gap. Because of these unusual properties a number of new effects appear in this material.
% and it makes scientific society study again even well known phenomena.\\
Thus, one of the examples of such unusual behavior are graphene plasmons.
They are collective oscillations of electronic density which can be found only in a spin singlet state. Currently, plasmons in graphene are under intensive investigation \cite{plasmontheory,plasmonNlayers}. There are a lot of works showing that dispersion law and properties of plasmons for Dirac electrons differ markedly from the plasmons in conventional 2D materials. For instance, in 2D semiconductors at long wavelength the plasma frequency $\omega_p\sim n^{\frac{1}{2}}$, whereas in graphene $\omega_p\sim n^{\frac{1}{4}}$. This is a direct consequence of the quantum relativistic nature of graphene \cite{plasmonA, plasmonC,plasmonD}. It is important to point out that recent experimental results \cite{plasmonB} agree well with theoretical predictions \cite{plasmonA}. Another significant feature of the graphene plasmons is long lifetime caused by the peculiar way of damping \cite{plasmon,plasmonlife}. Unlike conventional 2D materials, Landau damping occurs due to interband transitions in graphene. The edges of this region can be moved by manipulation of the doping level. As the authors of ref.\onlinecite{plasmonE} claim, for sufficiently large doping values low plasmon losses are possible in graphene.

Another type of excitation is the exciton, an electron and a hole coupled by their Coulomb attraction.
Exciton states in different semiconductors and $\pi$-bonded planar organic molecules have been studied for decades. Their energies lie in the band gap close to the bottom of conduction band. However, recently, it has become clear that they can also be found in semiconductor single walled carbon nanotubes (SWCNTs) \cite{excit1,excit,excitImplication} or in gapped graphene \cite{ribbon,tripextheory1, gapgraphene}. Low dimensionality of such structures results in strong Coulomb interactions. That opens the way for carbon-based optoelectronic devices operating at room temperature \cite{temp1,temp2}. A lot of experimental and theoretical investigations \cite{spataru} have been done in this area. Exciton states can be both spin - singlet or triplet. It is known that excitons in SWCNTs form a complex series of 16 exciton states. However, many authors maintain that only singlet exciton with odd parity is optically active and all the others are dark \cite{tripex3,tripex}. Therefore, there is a great deal less literature about the structure of low lying triplet excitons.
Despite of this, using different experimental techniques a satellite peak in photoluminescence spectrum attributed to the triplet dark exciton was measured independently by different groups \cite{tripex1,tripex4,tripex2}. It was discovered that magnetic excitons have longer lifetimes \cite{tripex2,time} than singlet ones, which makes them extremely important for photoelectronics applications and spin transport experiments \cite{magnon}.\\

During the last decade there was a discussion about the existence of a neutral spin triplet mode in a graphene sheet \cite{Graphite, resp1, resp2}. Some authors showed the existence of a spin-1 collective mode in undoped graphene using the Hubbard model and found its dispersion law \cite{Graphite}. However, in ref.\onlinecite{resp1} such solution was not found. Therefore, the issue remains open. In this article we prove that the triplet particle-hole excitations, magnons, do exist both in undoped and doped graphene.
%This problem is solved a great deal more accurate by solving the Schrodinger equation and taking into account all the matrix elements of the Hamiltonian.
This problem is solved taking into account all the matrix elements of the Hamiltonian.
In addition, we are interested in these excitations in the case of doped graphene. We carry out a full analysis of magnons spectrum properties what allows us to find conditions for existence of the spin-1 mode in the system. What is more, Coulomb interaction is considered in the context of the PPP-model.

The origin of the triplet excitations is very similar to the triplet excitons mentioned above or, e.g., to Stoner excitations observed in ferromagnets. Mainly, the existence of magnons in doped graphene is made possible by the gap in the two particle spectrum \cite{Graphite, graphene}. Indeed, the $e-h$ spectrum of graphene differs from the conventional 2D materials \cite{graphene}.
In metals, e.g., the $e-h$ continuum is formed by intraband transitions only. At the same time, in large band gap semiconductors %the situation is opposite
there is a room only for interband excitations because the chemical potential lies in the band gap. In graphene the situation is a bit more complicated. In undoped graphene the Fermi surface shrinks to the point where two cones of valence and conduction bands are connected. Therefore, there are only interband transitions. However, moving the chemical potential away from zero leads to appearance of intraband transitions lying in the low energy region.  Because of relativistic electronic properties of graphene these two processes form a window in the $e-h$ spectrum. The size of the gap depends on the value of doping. For direct transitions ($|\textbf{q}|=0$) the gap size equals to  $2\mu$.

We show that magnons are formed only in the presence of Coulomb interaction. Hence, the model describing correlation effects plays an important role in our calculations. We compare the results obtained with Hubbard and PPP models. It is shown that they differ, especially, when screening effects are not too strong. In addition, our calculations allow to get information about the effective mass and velocity of magnons.
\section{Model}
\begin{figure}[h!]
\includegraphics[scale=0.6]{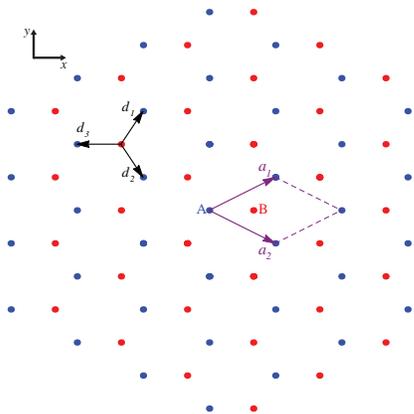}
\caption{Crystal structure of graphene. Two basis vectors
$\k a_1=\left\{\frac{\sqrt{3}a}{2},\frac{a}{2}\right\}$ and
$\k a_2=\left\{\frac{\sqrt{3}a}{2},-\frac{a}{2}\right\}$ form a unit cell. Blue and red circles indicate atoms belonging to the sublattices $A$ and $B$, respectively.
Each atom at site $B$ is connected to the three nearest neighbor atoms of type $A$ by the vectors: $\k d_1=\left\{\frac{a_{C-C}}{2},\frac{\sqrt{3}a_{C-C}}{2}\right\}$, $\k d_2=\left\{\frac{a_{C-C}}{2},-\frac{\sqrt{3}a_{C-C}}{2}\right\}$ and $\k d_3=\left\{-a_{C-C},0\right\}$. $a=\sqrt{3}a_{C-C}$ and $a_{C-C}\approx1.42{\AA}$ (distance between nearest neighbors). If $\k R^{(A)}_i$ is a position of atom $A$ in the unit cell $i$, then the position of $B$ atom in the same unit cell is: $\k R^{(B)}_i=\k R^{(A)}_i-\k d_3$. }
\label{lattice}
\end{figure}

It is known that graphene has a honeycomb lattice formed by two triangular sublattices i.e. $A$ and $B$, and a unit cell consists of two carbon atoms (Fig.\ref{lattice}).\\
We consider a Pariser-Parr-Pople (PPP) model which has successfully been used in different $\pi$ - conjugated systems \cite{PPP,PPP2}. The PPP Hamiltonian describing our system is written in the following way:
\begin{equation}
\label{ham}
\hat{H}=\hat{H}_{tb}+\hat{H}_{U}+\hat{H}_{V}.
\end{equation}
The tight-binding term is expressed like:
\begin{equation}
\label{tbham}
\hat{H}_{tb}=\sum\limits_{ij}\sum\limits_{\substack{\alpha_1\alpha_2\\ \sigma}}(t^{\alpha_1\alpha_2}_{ij}\hat{c}^{\dag}_{i\alpha_1\sigma}\hat{c}_{j\alpha_2\sigma}+h.c.),
\end{equation}
where operators $\hat{c}^{\dag}_{i\alpha\sigma}(\hat{c}_{i\alpha\sigma})$ create (annihilate) an electron with spin $\sigma$ ($\sigma=\uparrow,\downarrow$) in the unit cell $i$ on the atom belonging to the $\alpha$ ($\alpha=A/B$) sublattice. $t^{\alpha_1\alpha_2}_{ij}$ is the hopping amplitude between nearest neighbor sites.\\
The second term describes on-site Coulomb repulsion:
\begin{equation}
\label{hubham}
\hat{H}_{U}=U_0\sum_{i\alpha}\hat{n}_{i\alpha\uparrow}\hat{n}_{i\alpha\downarrow},
\end{equation}
where $U_0$ is the strength of on-site interaction and \ \ \ \ \ $\hat{n}_{i\alpha\sigma}=\hat{c}^{\dag}_{i\alpha\sigma}\hat{c}_{i\alpha\sigma}$.
\\
Finally, the last term is a long range interaction term defined by:
\def\MYsum{\mathop{
\lefteqn{\sum\nolimits^\prime} }}
\def\MYsum1{\mathop{
\lefteqn{\sum} }}
\begin{eqnarray}
\label{pppham}
\hat{H}_{V}=\sum\limits_{ij}\mathop{{\sum}^\prime}\limits_{\alpha_1\alpha_2}
V^{\alpha_1\alpha_2}_{ij}(\hat{n}_{i\alpha_1}-1)(\hat{n}_{j\alpha_2}-1),
%\hat{H}_{PPP}=
%\sum\limits_{i\neq j} \ \ V^{(AA)}_{ij}(\hat{n}^{(A)}_{i}-1)(\hat{n}^{(A)}_{j}-1)+
%\sum\limits_{i\neq j} \ \ V^{(BB)}_{ij}(\hat{n}^{(B)}_{i}-1)(\hat{n}^{(B)}_{j}-1)+\\
%\sum\limits_{ij} \ \ V^{(AB)}_{ij}(\hat{n}^{(A)}_{i}-1)(\hat{n}^{(B)}_{j}-1)+
%\sum\limits_{ij} \ \ V^{(BA)}_{ij}(\hat{n}^{(B)}_{i}-1)(\hat{n}^{(A)}_{j}-1),
\end{eqnarray}
where $\hat{n}_{i\alpha}=\sum\limits_{\sigma}\hat{c}^{\dag}_{i\alpha\sigma}\hat{c}_{i\alpha\sigma}$ is the number of electrons on the site $i\alpha$. $V_{ij}$ is the value of off-site Coulomb interaction. Prime in the second sum means that $\alpha_1\neq \alpha_2$ when $i=j$.\\
There are various ways to interpolate the long range part of the interaction. In this article we use the Ohno interpolation formula \cite{ohno}:
\begin{equation}
\label{ohno}
V^{\alpha_1\alpha_2}_{ij}=\displaystyle{\frac{U_0}{\sqrt{1+\frac{|\k R^{(\alpha_1)}_i-\k R^{(\alpha_2)}_j|^2}{a^2}}}},
\end{equation}
where $\textbf{R}_i^{(\alpha)}$ determines the position of $i\alpha$ atom and $a$ is a numeric parameter connected with screening effects.\\
Indeed, in a long range limit $V_{ij}=\frac{U_0a}{|\k R_i-\k R_j|}$. Wherefrom the relation between $a$ and dielectric constant of the substrate can be found:
\begin{equation}
\label{eps}
a=\frac{e^2}{4\pi\epsilon_0}\frac{1}{U_0\epsilon_r},
\end{equation}
where $e$ is an electron charge, $\epsilon_r$ is dielectric constant of a substrate and $\epsilon_0$ is vacuum permittivity.\\
From the expressions (\ref{pppham}) and (\ref{ohno}), obviously, that if $i=j$ and $\alpha_1=\alpha_2$, then $V_{ii}=U_0$ and we obtain the Hubbard term.\\
The main advantage of this approximation is that for specified value of $\epsilon_r$ there is only one parameter $U_0$ in the Hamiltonian what makes it much easier to analyze the properties of the system.

In this article the band structure of graphene is described by the tight-binding model, from which it is known that $\varepsilon(\k k)=\beta t |f(\k k)|$, where $f(\k k)=1+e^{\imath \k {ka}_1}+e^{\imath \k{ka}_2}$ and $\beta=+/-$ denotes either conduction or valence bands, respectively. The transformation between the real space and the momentum representations is:
\begin{equation}
c^{\dag}_{i\alpha\sigma}=\frac{1}{\sqrt{N}}\sum\limits_{\k k\beta}e^{-\imath \k {kR}_i}U^{-1}_{\k k\beta\alpha}c^{\dag}_{\k k\beta\sigma},
\label{transform}
\end{equation}
where $U^{-1}_{\k k\beta\alpha}$ are the matrix elements of the unitary matrix:
\begin{equation}
\hat{U}_{\k k}^{-1}=
\left(
  \begin{array}{cc}
    -1 & e^{-\imath\phi_{\k k}} \\
    e^{\imath\phi_{\k k}}  & 1 \\
  \end{array}
\right),
\end{equation}
and $\phi_{\k k}=\arg(f(\k k))$.\\
\begin{figure}[b]
\includegraphics[width=0.8\linewidth]{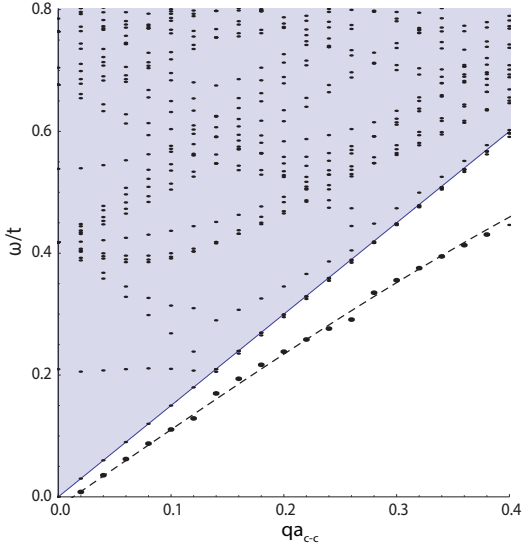}
\caption{Particle-hole spectrum and a spin-1 mode for undoped graphene. Blue area is a region of interband transitions in the absence of Coulomb interaction, while black dots denote the magnons spectrum computed using the Hubbard model ($\frac{U_0}{t}=4$). Dashed black line is a dispersion law of the magnons.}
\label{undoped}
\end{figure}
Because of a magnon being a bosonic excitation, it must be described by bosonic operators.
For these purposes, unlike ref.\onlinecite{iran}, we introduce three new operators creating a triplet $e-h$ pair with $S_z=1$, $S_z=0$, $S_z=-1$, respectively:
\begin{equation}
\begin{array}{c}
\Bdone 21=c_{2\usp}^{\dag}c_{1\dsp} \\
\Bdzero 21=\frac{1}{\sqrt{2}}\left(c_{2\usp}^{\dag}c_{1\usp}-c_{2\dsp}^{\dag}c_{1\dsp}\right)\\
\Bdmone 21=c_{2\dsp}^{\dag}c_{1\usp} \\
\end{array}
\end{equation}
and one operator which creates a singlet $e-h$ pair:
\begin{equation}
\Adzero 21=\frac{1}{\sqrt{2}}\left(c_{2\usp}^{\dag}c_{1\usp}+c_{2\dsp}^{\dag}c_{1\dsp}\right).\\
\end{equation}
In the definitions above $1$ and $2$ are general quantum numbers.
In general, these operators do not describe bosons. However, with respect to the ground state, the Fermi Sea state, and considering that the operator $c^{\dag}_{2\sigma}$ creates an electron with spin $\sigma$ above the chemical potential, $\mu$, while $c_{1\sigma}$ creates a hole with spin $-\sigma$ below the chemical potential, usual commutation relations are fulfilled:
\begin{equation}
\begin{array}{cc}
\langle FS|\left[\Bdone 21,\Bone {2'}{1'}\right]|FS\rangle=-\delta_{22'}\delta_{11'},\\
\langle FS|\left[\Bdmone 21,\Bmone {2'}{1'}\right]|FS\rangle=-\delta_{22'}\delta_{11'},\\
\langle FS|\left[\Bdzero 21,\Bzero {2'}{1'}\right]|FS\rangle=-\delta_{22'}\delta_{11'},\\
\langle FS|\left[\Adzero 21,\Azero {2'}{1'}\right]|FS\rangle=-\delta_{22'}\delta_{11'}.\\
\end{array}
\end{equation}
Therefore, it can be shown that the on-site interaction term of the Hamiltonian, $H_U$, can be presented in the invariant form:
\begin{equation}
H_U=\frac{U_0}{2N}\sum_{\substack{\k {21}\\ \k 1'\k 2'}}\mathfrak{T}^{\k 1\k 1'}_{\k {22'}}\left(-\bm{K}_{\k 2\k 2'}^{\dag}\cdot\bm{K}_{\k {11}'}+\frac{1}{2}\Adzero {\k 2}{\k 2'}\Azero {\k 1}{\k 1'}\right),
\label{hubb}
\end{equation}
where just for simplification:\\
$\k 2=\{\k k \beta_1\}$, $\k 1=\{\k k - \k q \beta_2\}$,$\k 1'=\{\k p-\k q \beta_3\}$, $\k 2'=\{\k p \beta_4\}$,
$\mathfrak{T}^{\k 1\k 1'}_{\k {22'}}=\sum\limits_{\alpha}U_{\k 1\alpha}(U_{\k 2\alpha})^*U_{\k 1'\alpha}(U_{\k 2'\alpha})^*$.\\
and
\begin{equation}
\k {K}^{\dag}_{ij}=
\left(
  \begin{array}{c}
    \displaystyle{\frac{\Bdone ij+ \Bdmone ij}{2}} \\
    \displaystyle{\frac{\Bdone ij- \Bdmone ij}{2\imath}}\\
   \frac{1}{\sqrt{2}}\Bdzero ij
  \end{array}
\right).
\end{equation}
The same procedure could be done and with $H_V$:
\begin{equation}
H_V=\frac{1}{N}\sum\limits_{\substack{\k {21}\\ \k 1'\k2'}}\mathfrak{F}_{\k {22'}}^{\k 1\k 1'}\left(-\k K_{\k {22}'}^{\dag}\cdot\k K_{\k {11}'}-\frac{1}{2}\Adzero {\k 2}{\k 2'}\Azero {\k 1}{\k 1'}\right),
\label{hvinv}
\end{equation}
where the following notations were introduced:\\
\begin{eqnarray*}
\mathfrak{F}_{\k {22'}}^{\k 1\k 1'}=\mathop{{\sum}^\prime}\limits_{\alpha_1\alpha_2}2V^{\alpha_1\alpha_2}_{\k q}U_{\k 1\alpha_1}(U_{\k 2\alpha_1})^*U_{\k 1'\alpha_2}(U_{\k 2'\alpha_2})^*,\\
%$\mathfrak{G}_{\k k\k p\k q}^{\k %\beta}=\mathop{{\sum}^\prime}\limits_{\alpha_1\alpha_2}V^{\alpha_1\alpha_2}_{\k %q}U_{\k k\alpha_1\beta_1}(U_{\k k-\k q\alpha_2\beta_2})^*U_{\k p-\k %q\alpha_2\beta_3}(U_{\k p\alpha_1\beta_4})^*$ and, finally, $V_{\k %q}^{\alpha_1\alpha_2}$ is the Fourier transform of (\ref{ohno}):
V_{\k q}^{\alpha_1\alpha_2}=\sum_jV^{\alpha_1\alpha_2}_{j}e^{-\imath\k q\k R_j}.
\end{eqnarray*}
%\begin{equation}
%\k {L}^{\dag}_{ij}=
%\left(
%  \begin{array}{c}
%    -\displaystyle{\frac{\Bdone ij+\imath \Bdzero ij}{\sqrt{2}}} \\
%   \Bdone ij \\
%    \displaystyle{\frac{\Bdone ij-\imath \Bdzero ij}{\sqrt{2}}}
%  \end{array}
%\right).
%\end{equation}
%Again, the minus before the second term describes attraction between an electron and a hole. $N_{\downarrow}$ is a total number of electrons with spin down.
\begin{figure}
    \includegraphics[width=1\linewidth]{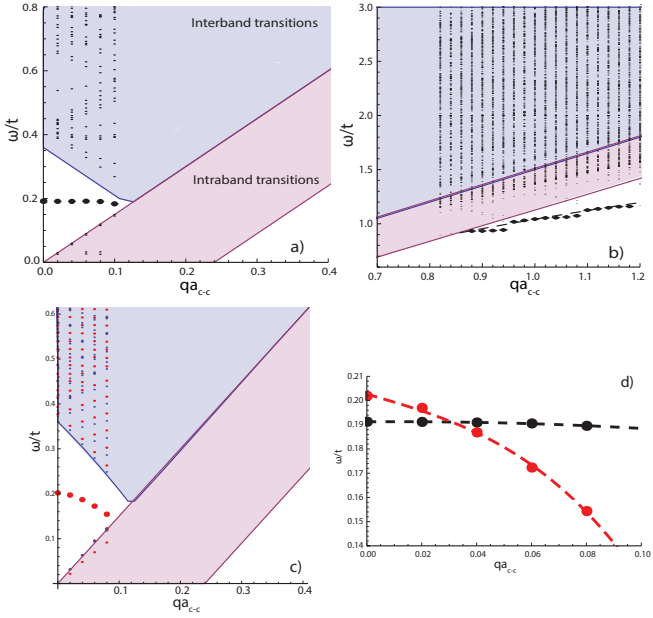}
\caption{Triplet excitations spectra calculated with the Hubbard model. a) and b) are low and high energy magnon modes in the doped graphene ($\frac{\mu}{t}=0.18$) for $\frac{U_0}{t}=4$, respectively. \\ c) Spectrum of magnons in the armchair (5,5) SWCNT for $\frac{U_0}{t}=3$ (red circles). Blue circles denote the spectrum for $\frac{U_0}{t}=0$. d) Magnons dispersion relations for graphene (black curve)  and for the (5,5) SWCNT (red curve). }
\label{plotHub}
\end{figure}
Now, let us consider the ground state of unexcited graphene is the Fermi Sea state $|FS\rangle$. In this article we are going to find the spectrum of $S_z=1$ excitations. It can be done without loss of generality because of the rotation invariance of the Hamiltonian. We look at a subspace of the total many-body Hilbert space which consists only of these excitations. Therefore, we specify the states living in that subspace as superposition of $e-h$ pairs:
\begin{equation}
\label{ansats}
|\psi_{\k q}\rangle^{(T)}=\sum\limits_{\k k \beta'\beta''}a_{\k k \beta'\beta''} \Bdone {\k k\beta'}{\k k -\k q \beta''} |FS\rangle,
\end{equation}
where $a_{\k k \beta'\beta''}$ are the coefficients which in general are complex.
Similarly, it is possible to construct a wave function describing singlet excitations of graphene corresponding to the plasmon mode:
\begin{equation}
\label{ansats2}
|\psi_{\k q}\rangle^{(S)}=\sum\limits_{\k k \beta'\beta''}a_{\k k \beta'\beta''} \Adzero {\k k\beta'}{\k k -\k q \beta''} |FS\rangle.
\end{equation}

Therefore, finding the spectrum of magnetic excitations reduces to solving the Schrodinger equation $H|\psi_{\k q}\rangle^{(T)}=\varepsilon_{\k q}|\psi_{\k q}\rangle^{(T)}$ for given values of $\k q$. By substitution of (\ref{ham}) and (\ref{ansats}) one gets a system for coefficients $a_{\k k\beta\beta'}$. In this article the spectrum is calculated for a piece of graphene composed of 1225 unit cells (or 2450 carbon atoms). We determined such size of the sample to eliminate the impact of size quantization effects. So, to obtain the spectrum of magnons the eigenvalues problem for the square ($1225\times1225$) matrix has to be solved.
\section{Results and Discussion}
\begin{figure}
    \includegraphics[width=0.8\linewidth]{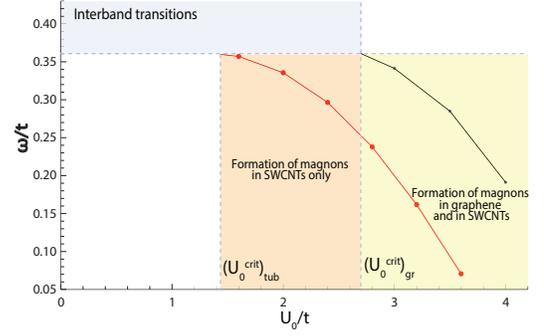}
\caption{Magnons energy dependence as a function of strength of Coulomb interaction for $|\k q|=0$ for graphene (black curve) and for (5,5) carbon nanotube (red curve).}
\label{DiffU}
\end{figure}
First of all, let us consider the case of undoped graphene sheet. Without Coulomb correlations the $e-h$ continuum consists only of the interband transitions. However, Coulomb interaction couples an electron and a hole what can lead to appearance of bound states. This situation is presented in the Fig.\ref{undoped}. There is a curve beneath the region of $e-h$ excitations, which corresponds to magnons. Thus, our calculation confirmed the Baskaran and Jafari proposal in the ref.\onlinecite{Graphite} on the existence of a magnetic collective mode.\\
%Furthermore, we found the dispersion law for new quasiparticles and it is close to the predicted analytical expression  $\hbar\omega\sim \alpha_1 q-\alpha_2 q^3$.

Now, if one dopes graphene, there is a window in the two particle spectrum and, as stated above, at small momenta and in the presence of Coulomb interaction we can expect to find new states in it.
In Fig.\ref{plotHub}a,b the spin-1 spectrum is presented. It was found numerically by solving the Schrodinger equation for Hubbard model ($V_{ij}=0$, $U_0=10.8eV$) and for different values of $\k q$. It is seen there are two new branches. The first one is in the long wavelength region (panel a)), while the second one, high energy branch, is located at small momenta (panel b)).
For better insight into the system behavior, it is useful to plot energy dependence of the bound state from the strength of Coulomb interaction for specific momentum value. The black curve in the Fig.\ref{DiffU} shows that energy of this bound state diminishes with rise of $U_0$. It is immediately seen from (\ref{hubb}). The triplet components of interaction give a negative contribution to the energy of the excited states, while a singlet one has a positive sign.
Therefore, under influence of Coulomb interaction low and high energy magnon modes are formed by the shift of the states from the interband and intraband transitions regions, respectively. The size of this shift is proportional to the value of $U_0$. Therefore, we can speak of a critical value $\left(U^{crit}_0\right)_{gr}$ above which the new states could be formed in the initially forbidden area. It is necessary to mention, that in ref.\onlinecite{RPAgr} authors also demonstrated the existence a magnon mode. However, it appears at much smaller values of interaction than in our calculations. Probably, such discrepancy is the consequence of considering all the matrix elements in the solution of the Schrodinger equation.\\
%Essentially, that even for $\k q=0$ the energy of triplet excitations is far from zero. It is in contrast to a plasmon state which arises from the low lying region of %intraband transitions and in the long wave limit its energy goes to zero.\\
\begin{figure}
\includegraphics[width=1\linewidth]{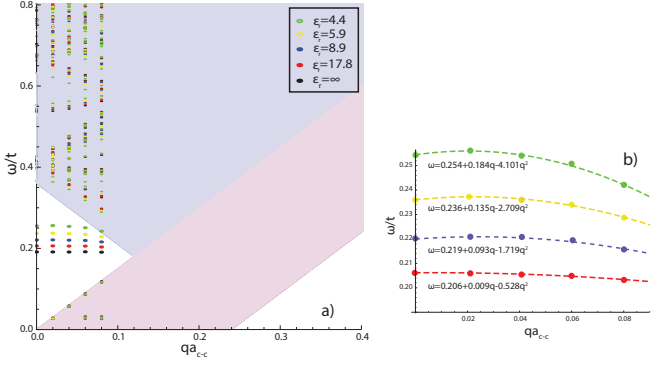}
\caption{a) PPP-spectrum of magnons in the doped graphene for different values of $\k q$ and $\epsilon_r$.
%Red dots correspond to $\epsilon_r=17.8$, blue to $\epsilon_r=8.9$, yellow to $\epsilon_r=5.9$ and green to $\epsilon_r=4.4$. Black dots denote the the spectrum calculated with Hubbard Hamiltonian.
b) Low energy magnons dispersion laws for different values of $\epsilon_r$.}
\label{plotPPP}
\end{figure}
\begin{figure}[b]
\includegraphics[scale=0.8]{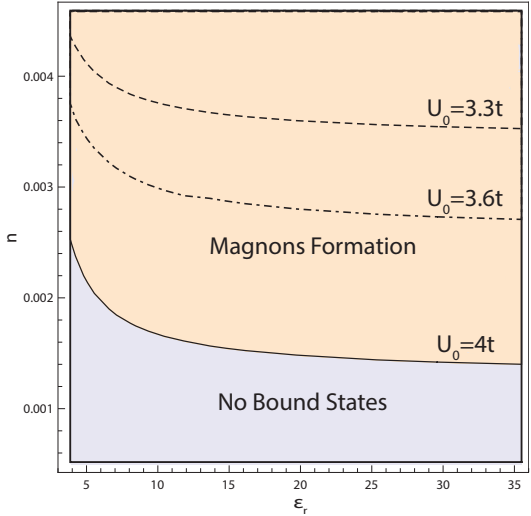}
\caption{The diapason of parameters $n$, $U_0$ and $\epsilon_r$ where magnons can exist. Solid line -- $\frac{U_0}{t}=4$; dashed -- $\frac{U_0}{t}=3.6$; dotdashed -- $\frac{U_0}{t}=3.3$. Above these edges a magnon mode can appear with given values of parameters.}
\label{phase}
\end{figure}
\begin{figure*}
\begin{center}
\begin{minipage}[h]{0.31\linewidth}
    \center{\includegraphics[width=0.9\linewidth]{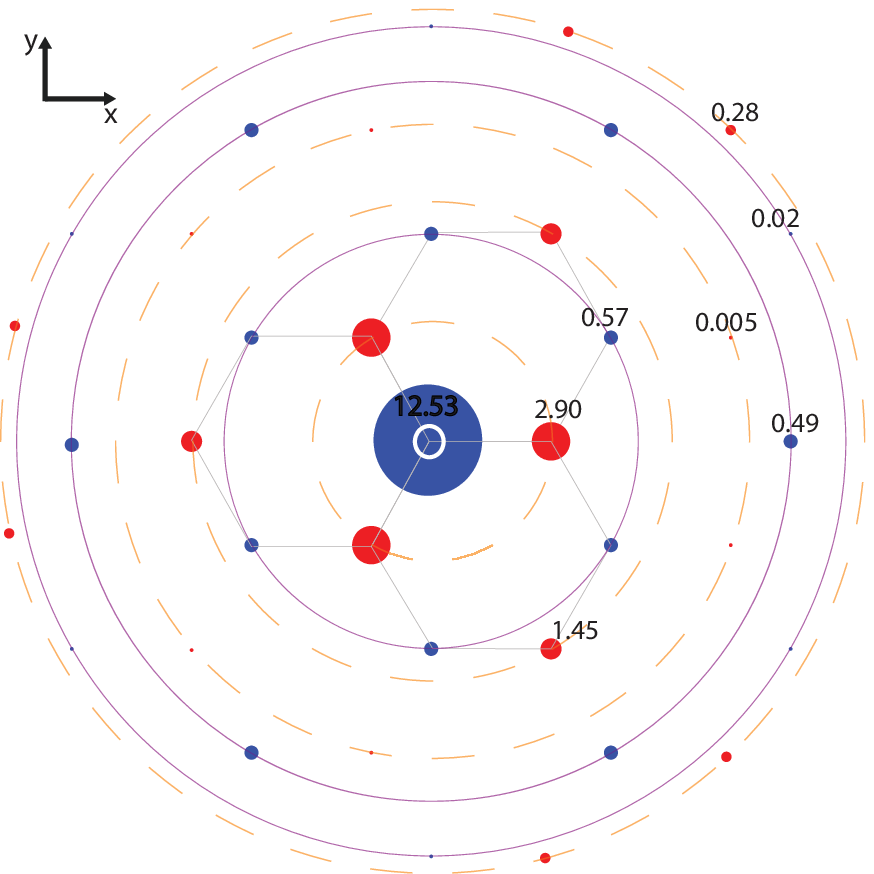}\\a)}
\end{minipage}
\hfill
\begin{minipage}[h]{0.31\linewidth}
   \center{ \includegraphics[width=0.9\linewidth]{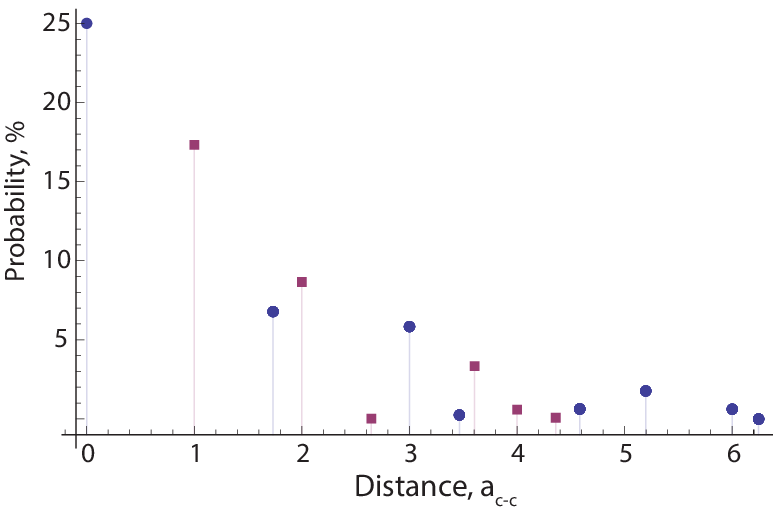}\\b)}
\end{minipage}
\hfill
\begin{minipage}[h]{0.31\linewidth}
    \center{\includegraphics[width=0.9\linewidth]{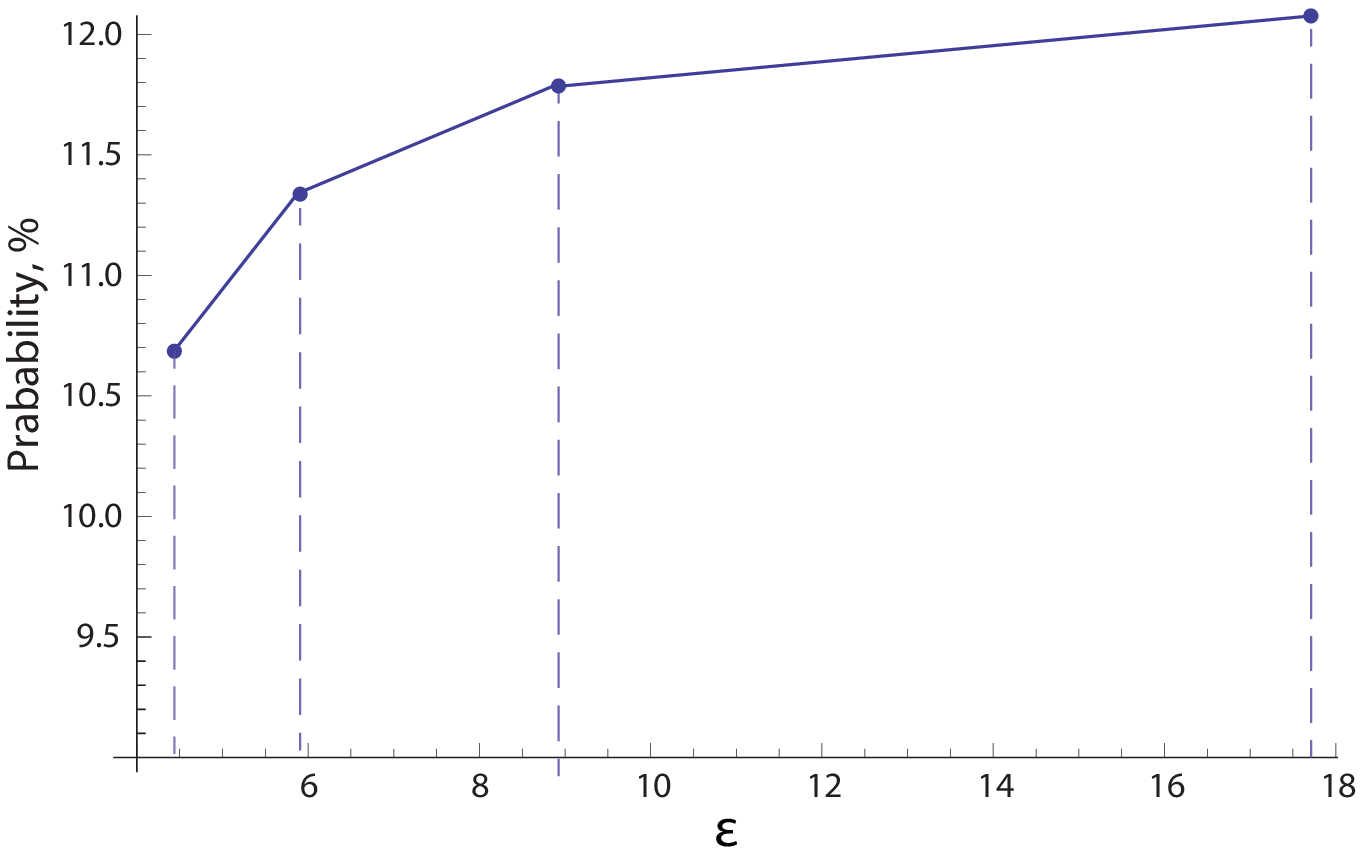}\\c)}
\end{minipage}
\end{center}
\caption{a) Spatial probability distribution of electron localization in graphene crystal for a fixed position of a hole calculated using Hubbard Hamiltonian ($\frac{U_0}{t}=4$). The radius of the circle is proportional to the probability of finding electron on the site. Blue/red circles are again atoms in sublattices $A/B$.  A hole is localized on the atom of type $A$ and is denoted by $\bigcirc$.
b)Full probability to find an $e-h$ pair in the graphene crystal as a function of distance between an electron and a hole positions.  Blue circles denote the case when both electron and hole are in the same sublattices. Pink squares describe the case when an electron and a hole belong to different sublattices. The probabilities were calculated for Hubbard model ($\frac{U_0}{t}=4$).
c)Probability dependence of localization both an electron and a hole on the same site as a function of $\epsilon_r$.}
\label{realplot}
\end{figure*}

Next, we consider carbon nanotubes. We would like to show that there are the same excitations in the metallic single-walled carbon nanotubes (mSWCNTs).
The only difference between this case and the one described above is different graphene and mSWCNT dispersion laws. \\

It is known that SWCNT consists of a graphene sheet that is rolled over a chiral vector $\k C_{nm}=n\k a_1+m \k a_2=(n,m)$, where $n$ and $m$ are some integers. There are three classes of the SWCNTs: armchair (n,n), zig-zag (n,0) and chiral (n,m). The condition for been metallic is $n-m=3q$, with integer $q$ \cite{tube1,tube2}.
In this article we consider only armchair nanotubes, but the calculations for zigzag one could be performed in the same manner. In the SWCNTs one of the component of the momentum is quantized, consequently, it is possible to show that the energy dispersion relation for armchair carbon nanotube is \cite{tube1}:
\begin{equation}
\varepsilon_{\nu}(k)=\sqrt{4 \cos \left(\frac{a k}{2}\right) \cos \left(\frac{\pi  \nu }{n}\right)+4 \cos ^2\left(\frac{a k}{2}\right)+1},
\end{equation}
where $k$ - is the continuous component of the wave vector, while $\nu$ corresponds to the discrete part of the wave vector (band index).
Considering $\nu=0$, we get the particle-hole spectrum similar to graphene. Therefore, it is instinctively clear that there should be a magnon mode, as well (Fig.\ref{plotHub}c,d).
However, there are a number of differences from the graphene magnons. The main one is lower value of critical Coulomb interaction. In Fig.\ref{DiffU} the red curve shows that the bound state appears in the armchair nanotube at $\left(U_0^{crit}\right)_{tub}\approx 1.4$, while in graphene $\left(U_0^{crit}\right)_{gr}$ is twice that.
%It is explained by the 1D nature of the nanotubes. In 1D because of the spatial confinement electrons fill stronger interaction.
Another sharp difference is that in the SWCNTs magnons have much stronger energy dependence from momentum than in graphene (Fig.\ref{plotHub}d). That is why, a magnon mode in the tubes is dumped at smaller values of the wave vector, comparing with the case of graphene. \\

Finally, we shall study the effect of long range Coulomb interaction.
Concentrating only on the low energy magnons, from Fig.\ref{plotPPP} it is seen that long range interaction shifts up the bound states from the their initial energy, calculated for $V_{ij}=0$. This shift is the smallest for large values of dielectric constant $\epsilon_r$. It is explained by (\ref{ohno}) from which we have $V_{ij}\sim\displaystyle{\frac{U_0}{\sqrt{1+\epsilon_r^2|\k R_{ij}|^2}}}$. For large $\epsilon_r$ the screening effects become stronger reducing the role of long range interactions.
The values of the dielectric constants of most semiconductors lie in the range $\epsilon_r\sim 10-16$. For such substrates, as can be seen, difference between the Hubbard model and the PPP-model is quite appreciable. Nevertheless, in a qualitative sense the structure of the spectrum is not changed. Therefore, substituting $U_0$ by some effective Coulomb interaction $V_{eff}$ the spin-1 spectrum can be calculated using the Hubbard model without loss of accuracy.\\
As shown above, there is a critical value of on-site Coulomb interaction below which the bound states are not formed. However, importantly that condition for appearance of these states depends not only on $U_0$, but also on $\epsilon_r$ and $\mu$ (Fig.\ref{phase}). One of the most essential features is growth of the value of doping needed for magnon formation when the strength of on-site interaction decreases.
%It is explained by the fact that rise of doping increases the size of the window what compensate "lack" of Coulomb interaction.
Taking $\epsilon_r=15$ and $U_0=4t=10.1eV$, the doping level corresponding to the bound state formation should be, roughly, $n=0.0018$ (or $\mu=0.27eV$) what is quite sensible and experimentally achievable value. However, for suspended graphene ($\epsilon_r=1$) the doping must be much higher.\\
%what, definitely, limits the experimental observation of magnons. So, choosing electrically opaque substrate is more likely because it demands much less doping.\\
%From the presented spectra it is not fully clear about electrical properties of magnons.
In Fig.\ref{plotPPP}b we present the magnons dispersion laws. As it is shown, the frequency dependence from momentum
%has quadratic form and it
becomes stronger when the role of $H_V$-term increases.  Approximating obtained data we can find the values of magnons velocity and effective mass. Thus, for the curve corresponded to $\epsilon_r=17.8$ we get that group velocity $v_{mag}=\frac{\partial\omega}{\partial q}\approx 6*10^3\frac{m}{s}$, what is three orders less then the Fermi velocity in graphene. Magnon effective mass is $m^*=\left(\frac{\partial^2\omega}{\partial q^2}\right)^{-1}=1.8 m_e$.\\

Finally, the last thing which is analyzed is how the magnons appear in the real space. To do it one has to calculate the probability to find an $e-h$ pair somewhere in the graphene lattice. For instance, if the hole position is fixed and it is localized on the atom of type $\alpha=A/B$, then the probability to find an electron in the position $\k R_i$ on the same type of atom is:
\begin{equation}
P_{\alpha\alpha}=\left|\frac{1}{2N}\sum\limits_{\k k}a_{\k k\beta\beta'}e^{-\imath \k k \k R_i}e^{-\imath\phi_{\k k}}\right|^2,
\end{equation}
and the probability to find an electron on the different type of atom is:
\begin{equation}
P_{\alpha_1\alpha_2}=\left|\frac{1}{2N}\sum\limits_{\k k}a_{\k k\beta\beta'}e^{-\imath \k k \k R_i}e^{-2\imath\phi_{\k k}}\right|^2.
\end{equation}
%Here $a_{\k k \beta\beta'}$ are the coefficients determined by solving the Schrodinger equation.
Evidently, we have $P_{AA}=P_{BB}$ and $P_{AB}=P_{BA}$.\\
Fig.\ref{realplot}a shows the spatial probability distribution of finding an electron somewhere in the crystal when the hole position is fixed (its position is marked on the graph as $\bigcirc$). One sees that the distribution function has the same symmetry as the graphene lattice. As we can see, there is the largest probability to find an electron on the same site as a hole what corresponds to a spin-flip. This situation is quite similar to that observed in a 1D chain of aligned spins when one spin-flip event causes a magnon.
As it is shown on Fig.\ref{realplot}b the probability distribution is exponentially decreasing as we move away from the position of a hole. This fact proves that magnons are really localized in the sample we used for our calculations and that the choice of its size was well founded.
Finally, it is quite interesting to know how long range interaction affects the probability of magnon observation (Fig.\ref{realplot}c). It shows that the probability that both an electron and a hole are on the same site increases with rise of $\epsilon_r$ and in the high $\epsilon_r$ limit it achieves the result obtained using the Hubbard model. Thus, it is shown that for large dielectric constants the magnons wave function is more localized in space.
%It is also essential result for experimental detecting of magnons that it is more preferable to have a substrate with a big dielectric constant.
\section{Conclusion}
In conclusion, in this article we investigat magnetic excitations in the doped graphene and in the armchair SWCNT. We show that a new mode appears in the gap of the particle-hole spectrum in the presence of Coulomb interactions in these systems. We argued that use of the Hubbard model with effective interaction gives correct results. Finally, it was shown that nanotube magnons are more sensitive to variation of the Coulomb interaction strength than graphene magnons.
%\begin{acknowledgments}

%\end{acknowledgments}

\nocite{*}

\bibliography{ReferencesForPRB_TripExitInGraph}% Produces the bibliography via BibTeX.

\end{document}